\documentclass[prd,aps,floats,preprint,preprintnumbers]{revtex4}
\usepackage{epsfig}
\newcommand{\be}{\begin{equation}}
\newcommand{\ee}{\end{equation}}
\newcommand{\bea}{\begin{eqnarray}}
\newcommand{\eea}{\end{eqnarray}}

\newcommand{\mn}{{\mu\nu}}
\begin{document}

\setlength{\unitlength}{1mm}
 
\title{Can we be tricked into thinking \\ that $w$ is less than $-1$?}

\author{Sean M. Carroll$^{1}$, Antonio De Felice$^{2}$ and Mark Trodden$^{2}$}

\affiliation{$^{1}$Enrico Fermi Institute, Department of Physics, and
Kavli Institute for Cosmological Physics, University of Chicago, 5640
S. Ellis Avenue, Chicago, IL 60637.
{\tt carroll@theory.uchicago.edu} \\
\\
$^2$Department of Physics, Syracuse University,
Syracuse, NY 13244-1130, USA.\\
{\tt defelice@physics.syr.edu, trodden@phy.syr.edu}}

\begin{abstract}
Dark energy candidates for which the equation-of-state parameter $w$
is less than $-1$ violate the dominant energy condition, and are
typically unstable.  In scalar-tensor theories of gravity, however,
the expansion of the universe can mimic the behavior of general relativity
with $w<-1$ dark energy, without violating any energy conditions.
We examine whether this possibility is phenomenologically viable by
studying Brans-Dicke models and
characterizing both the naturalness of the models themselves, and
additional observational constraints from limits on the
time-dependence of Newton's constant.  We find that only highly
contrived models would lead observers to measure $w<-1$.
\end{abstract}
\maketitle


\section{Introduction}
\label{intro}
The last several years have seen a sustained flow of cosmological data,
beginning with the observations of type Ia supernovae 
\cite{Riess:1998cb,Perlmutter:1998np,Garnavich:1998th}, bolstered by
large-scale redshift surveys (e.g.~\cite{Tegmark:2001jh}) and 
measurements of the cosmic microwave background (CMB) from 
the ground and from balloons~\cite{Ruhl:2002cz,Halverson:2001yy,Lee:2001yp,Pearson:2002tr}, 
culminating in the
exquisite full-sky maps of the WMAP satellite \cite{Spergel:2003cb}. 
These studies have indicated that the expansion of the universe is
accelerating (the scale factor obeys $\ddot{a} > 0$), and that the total
amount of clustered matter in the universe is insufficient to account
for the small value of its average spatial curvature.

Both of these features (acceleration and flatness) can be explained
in the context of conventional general relativity
by invoking a smooth, persistent dark-energy component, $X$ 
\cite{Wetterich:fm,Ratra:1987rm,Caldwell:1997ii,Armendariz-Picon:2000dh,Armendariz-Picon:2000ah,Mersini:2001su,Sahni:1999gb,Carroll:2000fy,Peebles:2002gy}. To be
compatible with the observed isotropy and homogeneity of our universe
on large scales, the energy-momentum tensor of the dark energy should
be that of a perfect fluid,
\begin{equation} 
\label{perfectfluid}
T_{\mu\nu}^{(X)} = (\rho_{X} + p_{X})U_{\mu} U_{\nu} + p_{X} g_{\mu\nu} \ ,
\end{equation} 
where $U^{\mu}$ is the fluid rest-frame four-velocity, $\rho_{X}$ is the
energy density and $p_{X}$ is the pressure.  The dark energy must
be smoothly distributed in order to escape detection in the dynamics
of gravitationally bound systems and large-scale structure.  To make
the universe accelerate, general relativity implies that
the pressure $p_X$ must be appreciable and negative.  From the Friedmann
equations
\be
  \label{f1}
  \left({\dot a \over a}\right)^2 \equiv H^2 = {8\pi G\over 3}\rho
  -{\kappa \over a^2}
\ee
and
\be
  \label{f2}
  {\ddot a\over a} = \dot{H} + 2H^2 = -{4\pi G\over 3}(\rho + 3p)\ ,
\ee
we find that acceleration only occurs if
the total pressure $p$ (dominated by the dark energy, since
matter is pressureless) is less than $-\rho/3$ (including both
matter and dark energy).

Since CMB observations imply that the spatial curvature is small
($|\kappa/a^2| \ll H^2$), it is convenient to set $\kappa=0$ and
consider a two-parameter set of cosmological models, characterized
by the density parameter in matter $\Omega_{m}\equiv 8\pi G\rho_{m}/3H^2$ 
and the dark energy equation-of-state parameter $w_X$,
\be
w_{X}=p_{X}/\rho_{X}\ .
\ee
Through the continuity equation
\be
  \dot{\rho}_X = -3H(1+w_X)\rho_X\ ,
\ee
the equation-of-state parameter governs the rate at which the dark
energy evolves as the universe expands,
\be
  {d\ln\rho_X \over d\ln a} = -3(1+w_X).
\ee
A strictly constant vacuum energy (a cosmological constant) would
have $w_X = -1$.  Observational constraints are often presented
as exclusion contours in the $\Omega_{m}$-$w_X$ plane.

Of course, it is impossible in principle to directly measure the
pressure in a component that is smoothly distributed, since there
are no pressure gradients (or at least no readily observable ones).
Constraints on $w_X$ actually derive from observations of the
behavior of the scale factor $a(t)$, and use the Friedmann equations
(\ref{f1},\ref{f2}) to translate these into limits on $w_X$.
Different observational methods will be sensitive to different
integrated behaviors of the dark energy density; it is nevertheless
useful to consider the instantaneous effective equation-of-state
parameter $w_{\rm eff}$ that would be derived from (\ref{f1},\ref{f2}) in a 
flat universe dominated by matter and dark energy:
\begin{equation}
\label{weff}
w_{\rm eff}=-(1+\alpha)\left(1+\frac23\,\frac{\dot H}{H^{2}}\right) \ ,
\end{equation}
where 
\begin{equation}
\label{alpha}
\alpha\equiv \frac{\Omega_{m}}{\Omega_{X}}
  = \frac{\Omega_{m}}{1-\Omega_{m}} \ 
\end{equation}
is the ratio of energy density in matter to that in dark energy.
Current observational
bounds~\cite{Hannestad:2002ur,Melchiorri:2002ux,Tonry:2003zg,Hannestad:2004cb} on
$w_{\rm eff}$ yield
\begin{equation}
\label{wbounds}
-1.48<w_{\rm eff}<-0.72
\end{equation}
at the 95\% confidence level.

While the task of identifying a compelling candidate source with equation of
state parameter in this region is a formidable challenge for fundamental
physics, the portion satisfying $w_{\rm eff}<-1$ is particularly troublesome
theoretically.  (Note that  recent data, e.g.~\cite{Riess:2004nr, Huterer:2004ch}, 
seem to suggest, at one or two sigma confidence, a $w_{\rm eff}$ changing with 
$z$ to values less than -1 today.) It is only possible to obtain $w<-1$ by violating the dominant
energy condition (DEC), which for a perfect fluid can be stated as \be \rho
\geq |p| \ .  \ee The physical motivation for the DEC is to prevent
instability of the vacuum or propagation of energy outside the light cone.
Nevertheless, such energy components have been known for some
time~\cite{Nilles:1983ge,Barrow:yc,Pollock:xe} and their role as possible dark
energy candidates was raised by Caldwell \cite{Caldwell:1999ew}, who referred
to DEC-violating sources as ``phantom'' components. The implications of these
have since been investigated by several authors (for some examples see
\cite{Sahni:1999gb,Parker:1999td,Chiba:1999ka,Boisseau:2000pr,Schulz:2001yx,Faraoni:2001tq,maor2,Onemli:2002hr,Torres:2002pe,Frampton:2002tu,McInnes:2002qw,Carroll:2003st,Caldwell:2003vq,Onemli:2004mb,Gonzalez-Diaz:2004vq,Cline:2003gs,Hsu:2004vr}.
Typically, DEC-violating sources with $w<-1$ are subject to violent
instabilities, although these may conceivably be cured in models with
higher-derivative kinetic terms~\cite{Carroll:2003st,Cline:2003gs,Hsu:2004vr}.

Despite the difficulties of model-building, it is certainly worthwhile
to consider the possibility that $w_{\rm eff}<-1$ when characterizing
observational constraints, if only to keep open the possibility of
a surprising discovery.  In addition,
given how little we understand about dark energy, we should keep
an open mind about the true explanation for the acceleration of
the universe.  One alternative that has been explored is a modification
of general relativity that would only become important on
cosmological scales \cite{Carroll:2003wy,Capozziello:2003tk,Freese:2002sq,Vollick:2003aw,Soussa:2003re,Nojiri:2003ni,Elizalde:2004mq,Flanagan:2003iw,Arkani-Hamed:2003uy,Gabadadze:2003ck,Dvali:2004ph,Moffat:2004nw}.  If the
Friedmann equations (\ref{f1},\ref{f2}) are not valid, the 
reconstruction (\ref{weff}) of the equation-of-state parameter
would similarly be invalid.  Therefore, it becomes conceivable
that we could measure the effective value
$w_{\rm eff}$ (as defined by (\ref{weff})
to be less than $-1$, even if the actual dark energy source obeys
the DEC, or if there is indeed no dark energy at all.  In other
words, we could be tricked into thinking that $w_X<-1$ by assuming
the validity of general relativity (see also~\cite{Stefancic:2003bj} for examples of
non-phantom matter behaving as phantom matter).

>From (\ref{weff}) it is clear that we infer $w_{\rm eff}<-1$ today if
\begin{equation}
\label{criterion}
\frac{\dot H_{0}}{H^{2}_{0}}>-\frac32\,\frac{\alpha}{1+\alpha} \ ,
\end{equation}
where the subscript $0$ indicates present-day values of
quantities. 

In this paper we consider a simple class of modifications of general
relativity: scalar-tensor theories, featuring a scalar field $\phi$ that
interacts non-minimally with gravitation ({\it e.g.}, through a direct
coupling to curvature).  A wide variety of alternatives to GR can be cast as
scalar-tensor theories, at least in some range of validity.  Focusing on
Brans-Dicke (BD) theories~\cite{Brans:1961sx}, we examine whether a
scalar-tensor theory of gravity could account for the acceleration of the
universe and yield $w_{\rm eff}<-1$, while remaining consistent with other
experimental constraints.  In particular, the cosmological evolution of the
scalar $\phi$ leads to time-dependence of Newton's gravitational constant $G$,
which are constrained by solar-system tests of gravity.  We find that only
extremely unnatural and contrived models lead to an inference of $w<-1$, even
in this wide class of extensions to general relativity.


\section{Brans-Dicke theories}
A particular class of theories of gravity beyond General Relativity 
are Brans-Dicke theories~\cite{Brans:1961sx} with a potential,
which may arise, for example, from the dimensional reduction of a
higher dimensional theory. These theories consist of a metric and a
Brans-Dicke scalar field $\varphi$, with action
\begin{equation}
\label{BDaction}
S_{\rm BD}=\int d^4x\, \sqrt{-g}\left[\varphi R
-\frac{\omega}{\varphi}\left(\partial_{\mu}\varphi\right)
\partial^{\mu}\varphi -2V(\varphi)\right] +\int d^4x\, 
\sqrt{-g}{\cal L}_M(\psi_i,g) \ ,
\end{equation}
where ${\cal L}_M(\psi_i,g)$ is the Lagrangian for matter fields $\psi_i$ and
$\omega$ is a constant. In this frame, the Jordan frame, matter is
minimally-coupled to gravity and hence test particles fall freely along
geodesics of the metric $g_{\mu\nu}$. Scalar-tensor theories such as these have
been considered previously as a way to solve the the fine-tuning problems of
quintessence (see~\cite{Perrotta:1999am,Baccigalupi:2000je,Matarrese:2004xa}.)

The predictions of Brans-Dicke theories differ from those of GR due to the
presence of a new scalar component to gravity. Since GR is well-tested in the
solar system, these deviations must be smaller than the accuracy of current
observations. This can happen in one of two ways. Either the potential is
sufficiently confining to render the Brans-Dicke scalar essentially constant
in the solar system or, in the presence of a sufficiently weak potential, the
parameter $\omega$ must satisfy the bound $\omega>40000$, obtained using
signal timing from the Cassini spacecraft~\cite{Cassini}. However, this bound may be 
weaker on cosmological scales than in the solar system~\cite{Clifton:2004st}.

We will be interested in two possibilities. First, that a smoothly distributed
background matter component, described by ${\cal L}_M$ and minimally coupled
to the Brans-Dicke sector, may fuel faster than exponential expansion. Second,
that the Brans-Dicke scalar itself may lead to a similar effect, without
appealing to sources outside the gravitational sector of the theory.

It is convenient to perform both a conformal transformation and a
field redefinition to obtain an Einstein frame description of the
theory. 

We define a canonically-normalized version of the Brans-Dicke scalar 
via
\be
e^{\sigma/\sigma_*} = 16\pi G\phi\ ,
\ee
where
\begin{equation}
\label{sigmastar}
\sigma_* \equiv \sqrt{\frac{2\omega +3}{16\pi G}} \ ,
\end{equation}
with $\omega$ the Brans-Dicke parameter.  
Denoting the Jordan-frame metric by $g_\mn$, the Einstein-frame metric is
\begin{equation}
\label{conftrans}
{\bar g}_{\mu\nu}\equiv e^{\sigma/\sigma_*}g_{\mu\nu}
\end{equation}
and the resulting action becomes
\begin{equation}
\label{Einsteinaction}
S_{\rm E}=\int d^4x\, \sqrt{-{\bar g}}\left[\frac{1}{16\pi G} {\bar
R}-\frac{1}{2}\left(\partial_{\mu}\sigma\right)\partial^{\mu}\sigma
-U(\sigma)\right] +\int d^4x\, \sqrt{-{\bar g}}{\cal L}_M(\psi_i,{\bar
g},\sigma) \ ,
\end{equation}
where
\begin{equation}
\label{potential}
U(\sigma)\equiv 2 e^{-2\sigma/\sigma_*}V[\varphi(\sigma)] \ .
\end{equation}
This description now has the advantage that the gravitational sector is of
Einstein form, with a minimally coupled scalar field, but the disadvantage
that the matter does not freely-fall along the geodesics of the Einstein-frame
metric.

Further, the relationship between the energy-momentum tensors in the
two frames is
\begin{equation}
T^{\mu\nu}=\Omega^{3}\,\bar T^{\mu\nu} \ .
\end{equation}
Therefore, if the energy conditions are not violated in the Einstein
frame they will not be violated in the Jordan frame, since $\Omega$ is
strictly positive.  

Let us now study the Einstein-frame equations
of motion. We define the Einstein-frame scale
factor and time coordinate by
\bea
  \bar{a} &=& e^{\sigma/2\sigma_*}a \cr
  d\bar{t} &=& e^{\sigma/2\sigma_*} dt 
\eea
and let a prime denote differentiation with respect to $\bar{t}$, so
that the Einstein-frame Hubble parameter is
\be
  \bar{H} \equiv {\bar{a}' \over \bar{a}}\ .
\ee

The Friedmann equations are then
\begin{equation}
{\bar H}^2=\frac{8\pi G}3\left(\rho_{\sigma}+ \bar{\rho}\right) \ ,
\label{effriedmann}
\end{equation}
\begin{equation}
\bar H'=-4\pi G\left[(\sigma')^2+\bar{\rho}+\bar{p} \right] \ ,
\label{hbarprime}
\end{equation}
and the equation of motion for $\sigma$ is
\begin{equation}
\sigma''+3\,\bar H\,\sigma'+U_{,\sigma}={1\over 2\sigma_*}
\left(\bar{\rho}-3\bar{p}\right)\ .
\label{efscalar}
\end{equation}
Here we have also defined $\rho=e^{2\sigma/\sigma_*}\bar\rho$ and 
$p=e^{2\sigma/\sigma_*}{\bar p}$, where $\rho$ and $p$ are respectively
the Jordan frame energy density and pressure of matter. Since
$\sigma$ is a canonically-normalized, minimally-coupled scalar field,
its energy density and pressure are given by the usual definitions
\begin{eqnarray}
\rho_{\sigma} & = & \frac{1}{2}\sigma'^2 + U(\sigma) \ ,\\
p_{\sigma} & = &   \frac{1}{2}\sigma'^2 - U(\sigma) \ .
\end{eqnarray}

Because cosmological observations, and in particular those quantities
entering~(\ref{criterion}), involve Jordan frame quantities, we need
to know how to transform between our easily-interpretable
Einstein-frame quantities and those in the Jordan frame. This is achieved by
the following expressions \be
\label{hubbletransform}
H=e^{\sigma/2\sigma_*}\left(\bar H-\frac{\sigma'}{2\sigma_{*}}\right) \ ,
\ee
and
\be
\label{hdottransform}
  \dot{H} = e^{\sigma/\sigma_*}\left[ 
  \bar{H}' -\frac{\sigma''}{2\sigma_{*}} +
  {\sigma' \over 2\sigma_*}
  \left(\bar H-\frac{\sigma'}{2\sigma_{*}}\right)\right]\ .
\ee

Substituting the Einstein-frame equations of motion into the expressions
(\ref{hdottransform}) and (\ref{hubbletransform})
we obtain, after some algebra
\bea
  \dot{H} &=& e^{\sigma/\sigma_{*}} \left\{-{8\pi G\over (2\omega+3)}
  \left[(\omega+2)\left(\bar\rho + (\sigma')^2\right) 
  +\omega\bar{p}\right] 
  \right. \cr
  && \qquad \qquad
  + \left. \left({4\pi G\over 2\omega+3}\right)^{1/2}\left[
  4\sigma'\sqrt{{8\pi G\over 3}(\bar\rho + \rho_\sigma)}
  +U_{,\sigma}\right] \right\}
  \label{hdot2}
\eea
and
\be
  H = e^{\sigma/2\sigma_{*}} \sqrt{4\pi G}\left[\sqrt{2\over 3}
  \sqrt{\bar\rho + {1\over 2}(\sigma')^2 + U} -\sqrt{{(\sigma')^2
  \over 2\omega + 3}}\right]\ .
  \label{h2}
\ee
These are the fundamental expressions relating the observable quantities
$H$ and $\dot H$ to sources as measured in the Einstein frame.

\section{A Perturbative Approach}
\subsection{Acceleration from a Dark Energy Component in Brans-Dicke Theories}
Let us now focus on the case in which the source of cosmic acceleration is some
dark-energy component in the matter sector (rather than the BD field itself).
Since we are interested in the present day, we have $\sigma=\sigma_0 = 0$. 
We assume that the BD field provides a negligible contribution to the evolution of the 
universe, so that
\be
  (\sigma')^2 ,\ U(\sigma)\ \ll \bar\rho \ ,
\ee
and that the theory is consistent with solar system test of gravity, 
requiring $\omega \gg 1$.  However, we will keep quantities that are first-order 
in $1/\sqrt{\omega}$, since these are
important for our effect.  The ``matter'' sector will consist of ordinary matter
plus the dark-energy component $X$.

Expanding and evaluating (\ref{hdot2}) and (\ref{h2}) at the present day,
after some algebra we obtain
\be
  \dot{H}_0 \approx -4\pi G\left(1+ {w_X\over 1+\alpha} - 4\xi-\eta\right)
  \bar\rho_0\ ,
\ee
\be
  H_0^2 \approx 4\pi G\left({2\over 3} - \xi\right)
  \bar\rho_0\ ,
\ee
where we have introduced the parameters
\be
  \xi \equiv \left[{2(\sigma_0')^2\over 3(2\omega+3)\bar\rho_0}\right]^{1/2}
  \label{xidef}
\ee
and
\be
  \eta \equiv {2U_{,\sigma 0}\over \sqrt{16\pi G(2\omega+3)}\bar\rho_0}\ .
  \label{etadef}
\ee
(Note that, since $\sigma_0=0$, we have $\bar\rho_0 = \rho_{X0}+\rho_{m0}$,
so we could drop the tildes, but we won't bother.)
We thus have
\be
  {\dot{H}_0 \over H_0^2} \approx - 3 \left({1+{w_X\over 1+\alpha} - 4\xi - \eta
  \over 2 - 3\xi}\right)\ ,
\ee
which implies
\be
  w_{\rm eff} \approx {2w_X - (5\xi + 2\eta)(1+\alpha) \over 2-3\xi}\ .
  \label{weff1}
\ee

So far we have not assumed that the parameters $\xi$ and $\eta$ are small.  However, 
note that
\be
  \xi^2 = {2(\sigma_0')^2 \over 3(2\omega+3)\bar\rho_0}\ .
\ee
Since we have already assumed that $(\sigma_0')^2 \ll \bar\rho_0$, and also
that $\omega \gg 1$, we see that $\xi$ is certainly very small.  It is therefore
legitimate to rewrite (\ref{weff1}) as
\be
  w_{\rm eff} \approx w_X - \left(1+{5\over 2}\alpha\right)\xi
  - (1+\alpha)\eta\ .
  \label{weff2}
\ee
Note that for $\alpha = 3/7$, this is
\be
  w_{\rm eff} \approx w_X - \frac{29}{14}\xi
  - {10\over 7}\eta\ \ ,
  \label{weff3}
\ee
and since $\xi$ is extremely small (less than $10^{-3}$), the second
term on the right hand side will have a negligible effect on $w_{\rm eff}$.

Now let us turn to $\eta$.  If we Taylor-expand the potential to first order
about the present day value $\sigma=0$,
\be
\label{taylorpot}
  U(\sigma)\approx U_{0}\left(1+\lambda\,\frac\sigma{\sigma_{*}}\right) \ , 
\ee 
where
$U(0)\equiv U_{0}$ and
$U_{,\sigma}(0)\equiv\lambda\,U_{0}/\sigma_{*}$,
we obtain
\be
\eta \approx {2\lambda U_0 \over (2\omega+3)\bar\rho_0}\ .
\ee
Again, since we have assumed $U_0 \ll \bar\rho_0$ and $\omega\gg 1$,
this looks very small except for the freedom associated with the dimensionless
parameter $\lambda$.  Thus, it is possible to obtain $w_{\rm eff}$ detectably
below $-1$ if
\be
  \lambda \geq \omega\left({\bar\rho_0\over U_0}\right)\ .
\ee

Thus, we certainly can get $w_{\rm eff}< -1$ with matter sources that obey the 
Dominant Energy Condition.  However, we should think about what it means to have
$\eta$ be large while $\xi$ remains small.  From the definitions (\ref{xidef})
and (\ref{etadef}), this occurs only if the time derivative of the BD field is
small while the slope of its potential is very large.  This will only happen
for very finely-tuned conditions -- either a very sudden change in the slope
of the potential, or when the initial conditions for the field are chosen such
that it has recently been climbing {\sl up} the potential and is near a local
maximum today.  Hence, although it is possible to obtain $w_{\rm eff}< -1$ 
behavior in BD theories, it is by no means natural.

\subsection{Acceleration from the Brans-Dicke Scalar with no Dark Energy}
The right hand side of the Einstein-frame Friedmann equations are sourced not 
only by matter, but also by the BD scalar $\sigma$. It is natural to ask whether 
super-exponential acceleration can be obtained purely from this modification of general
relativity.

Making the approximation~(\ref{taylorpot}), using~(\ref{hubbletransform}),
evaluated today and defining $x\equiv\sigma'_{0}/\sqrt{\rho_{cr,0}}$
we obtain \bea \frac{\dot
H_{0}}{H_{0}^{2}}&=&\frac32\,\frac{2w+4}{2w+3}\,x
\left[\frac4{\sqrt{3}}\,\frac{\sqrt{2w+3}}{2w+4}\,
\sqrt{x^{2}+\frac{2U_{0}}{\rho_{cr,0}}+
\frac{2\alpha}{1+\alpha}}-{}\right. \nonumber\\ &&\left.{}-x\right]+
\frac{U_0}{\rho_{cr,0}}\,\frac{3\lambda}{2w+3}- 3\,
\frac{w+2}{2w+3}\,\frac\alpha{1+\alpha}\ .  
\eea
The Friedmann equation can then be rewritten as
\be
\frac{U_{0}}{\rho_{cr,0}}=\left[1+\sqrt{\frac3{4w+6}}\,x\right]^{2}
-\frac{x^2}2-\frac\alpha{1+\alpha}\ ,
\label{U0}
\ee
which allows us to obtain
\be
\frac{\dot H_{0}}{H_{0}^{2}}=\frac1{2w+3}\left\{
\frac32\,x\left[
\frac4{\sqrt{3}}\,\sqrt{4w+6}-2wx\right]+3\lambda\,\frac{U_{0}}{\rho_{cr,o}}-3\,\frac\alpha{1+\alpha}\,(w+2)\right\}\ .
\label{plotH}
\ee

{}From equation~(\ref{U0}) we can see that in the limit $|x|\ll 1$, which
we may term the {\it slow-roll} regime, we obtain
\be
\frac{U_{0}}{\rho_{cr,0}}\approx\frac1{1+\alpha}\ ,
\ee
which, from~(\ref{plotH}) yields
\bea
w_{\rm eff}<-1&\quad{\rm for}\quad&
\lambda>\frac\alpha2\label{lamb}\\
\dot H_{0}>0&\quad{\rm for}\quad&
\lambda>\alpha\,(\omega+2)\nonumber
\eea
Therefore, given that $\omega$ is typically required to be much larger
than unity, there exists a range in which the universe is
not superaccelerating, but in which we infer $w_{\rm eff}<-1$, 
even without an additional dark energy component.  Again, though, we
require a very large value of the parameter $\lambda$ characterizing
the slope of the potential.

\section{Constraints from the TIme-Variation of Newton's Constant}
Thus far we have focused only on the behavior of the effective
equation of state parameter. However, in Brans-Dicke theories it is
important to remember that, for a weak BD potential, the field
$\varphi$ plays the role of a dynamical Newton constant $G_{\rm eff}$
via \be 16\pi G_{\rm eff} = \frac1\phi\,\frac{2w+4}{2w+3} \ .
\label{effectiveG}
\ee
Experimental constraints on the time variation of Newton's constant yield
\be 
\frac{|\dot G_{\rm eff,0}|}{G_{\rm eff,0}}<6\times10^{-12}\ {\rm yr}^{-1}\ .
\ee
Since~(\ref{effectiveG}) gives
\be
\frac{|\dot G_{\rm eff}|}{G_{\rm eff}}=\Omega^{1/2}\,\frac{|\phi'|}{\phi}=
\Omega^{1/2}\,\frac{|\sigma'|}{\sigma_*}
\label{G}
\ee
this then implies 
\be
\frac{|\sigma'_0|}{\sigma_*}<6\times10^{-12}\ {\rm yr}^{-1}\ ,
\ee
or
\be
|x| = \sqrt{\frac{2w+3}6}\, \frac1{H_0}\, \frac{|\dot G_{\rm eff,0}|}{G_{\rm eff,0}}
<\sqrt{\frac{2w+3}6}\,\frac{5.88\times10^{-2}}h\ ,
\label{solar}
\ee
where we have written $H_0=100h$ km s$^{-1}$Mpc$^{-1}$, with $h=0.72\pm0.08$.

We would like to know what range of $w_{\rm eff}$ is possible today
while remaining consistent with bounds on the time-variation of
Newton's constant. To this end it is convenient to define 
\be
y=\frac1{H_0}\,\frac{\dot G_{\rm eff,0}}{G_{\rm eff,0}}\ .  
\ee 
In
terms of this variable equation~(\ref{U0}) becomes
\be
\frac{U_0}{\rho_{cr,0}}\equiv \gamma(y)= \left(1+\frac
y2\right)^2-\frac{2w+3}{12}\,y^2-\frac\alpha{1+\alpha}\ , 
\ee 
and
using equation~(\ref{plotH}) we obtain 
\be w_{\rm eff} =
-(1+\alpha)\left[1+\frac13\,y\,(4-wy)+
\frac{2\lambda}{2w+3}\,\gamma(y)-\frac{2\alpha}{1+\alpha}\,
\frac{w+2}{2w+3}\right] \ .  
\ee

\begin{figure}[ht]
\centering
\includegraphics[width=3.5in]{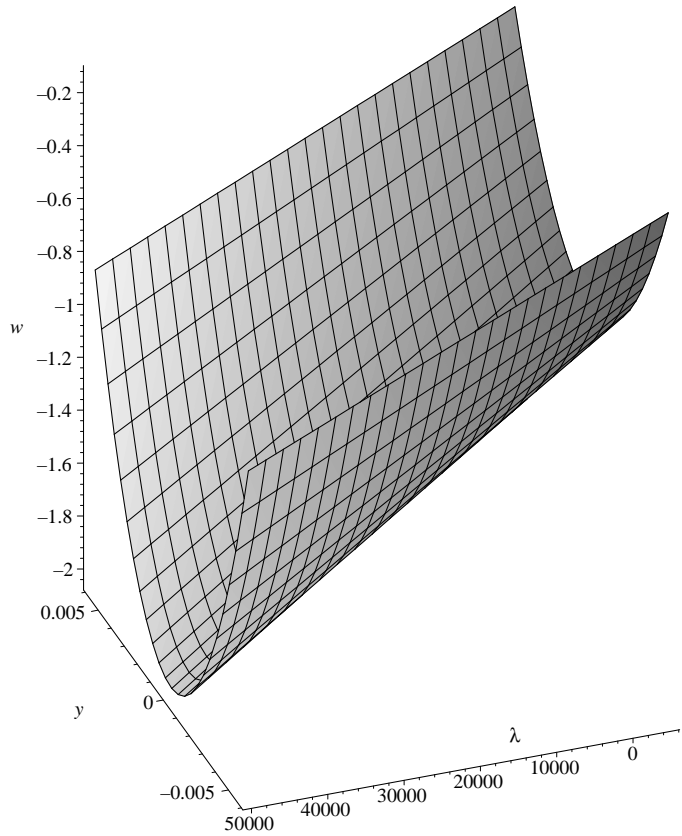}
\caption{$w_{\rm eff}$ as a function of $y\equiv \frac1{H_0}\,
\frac{\dot G_{\rm eff}}{G_{\rm eff}}$ and $\lambda$.}
\label{fig1}
\end{figure}

Clearly, $w_{\rm eff}$ is a function of three parameters:
$\Omega_m/\Omega_{X}\equiv\alpha=3/7$, $\lambda$, giving the slope of the
potential today, and the Brans-Dicke parameter $\omega$.  For fixed
$\omega = 40000$, we plot this expression as a function of
the variables $y$ and $\lambda$ in fig.~\ref{fig1}. As can be seen, negative
values of $\lambda$ correspond to positive $w_{\rm eff}$.

\begin{figure}[ht]
\centering
\includegraphics[width=2.5in]{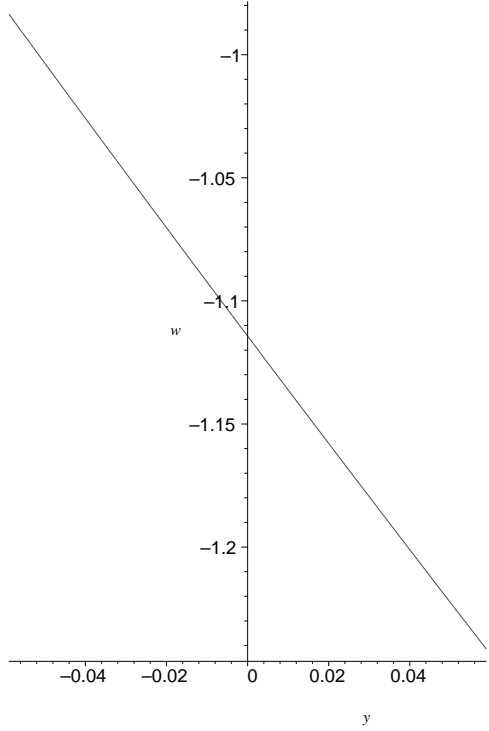}
\caption{$w_{\rm eff}$ as a function of $y\equiv \frac1{H_0}\,
\frac{\dot G_{\rm eff}}{G_{\rm eff}}$ with $\omega=1$.}
\label{fig2}
\end{figure}

\begin{figure}[ht]
\centering
\includegraphics[width=2.5in]{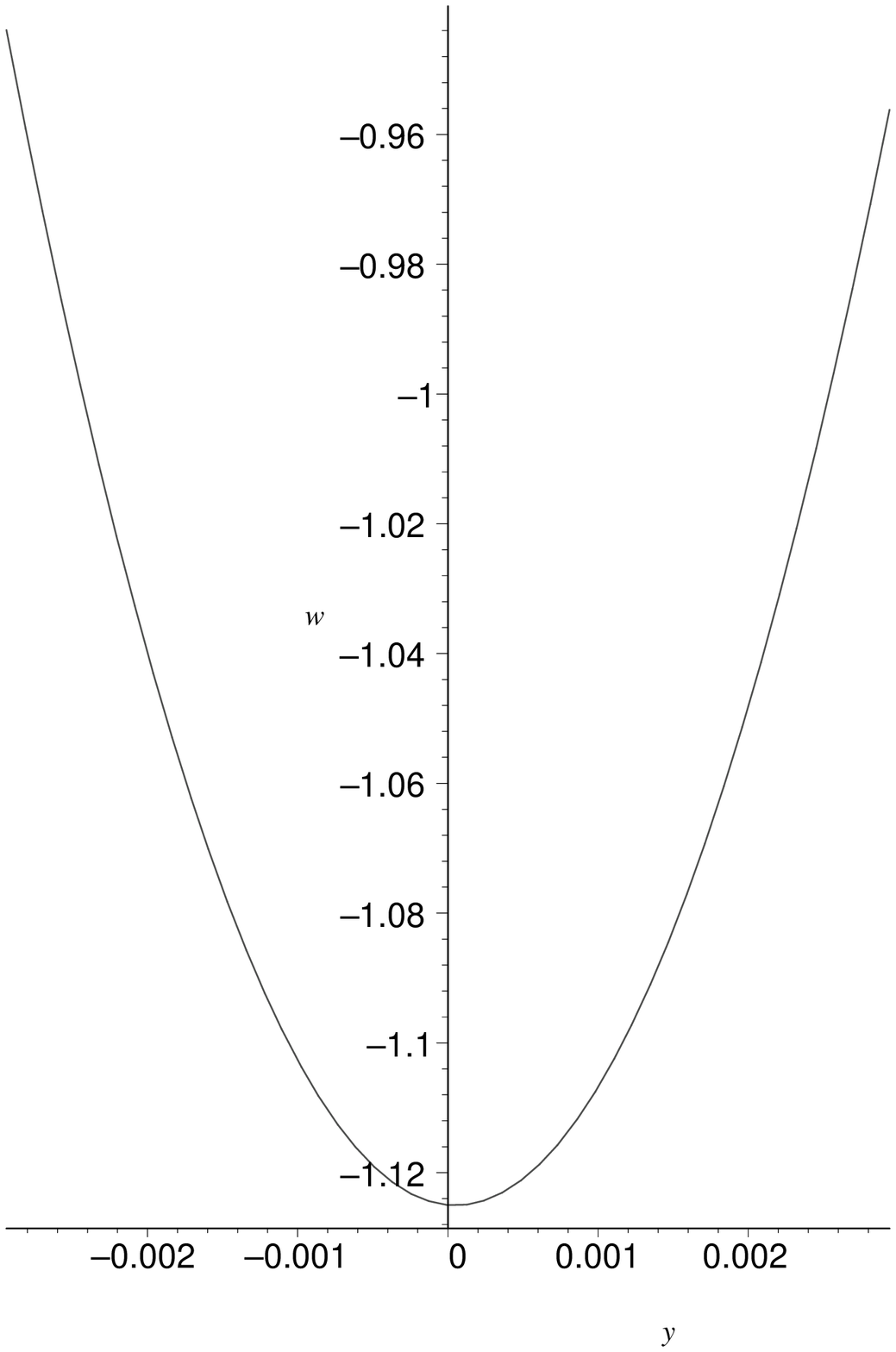}
\caption{$w_{\rm eff}$ as a function of $y\equiv \frac1{H_0}\,
\frac{\dot G_{\rm eff}}{G_{\rm eff}}$ with $\omega=40000$ and 
$\lambda\gg\alpha$.}
\label{fig3}
\end{figure}

Fixing the parameters $\lambda$ and $\omega$ and plotting this function
for $|y|<5.88\times10^{-2}$, we see that, in order to get values of $w_{\rm
eff}\approx-1.1$, we need $\lambda>0.8$ for $\omega=1$ (see fig.~\ref{fig2}),
or $\lambda>5000$ for $\omega=40000$ (see fig.~\ref{fig3}). In this last case we
must also demand $|y|<0.001$.

{}From the graphs we see that in order to obtain $w_{\rm
eff}\sim-1.1$ with $\omega=1$ we require $\lambda\sim\alpha$.  However, this condition
implies a severe fine-tuning. The equation for the scalar field
requires that the second derivative be negative and much larger in
magnitude than the first derivative.  Therefore, $\sigma$ must be
close to a maximum today. This is consistent with the toy potential we
used as a test. 

\section{Study of general solutions for BD acceleration}
To establish the main result of this paper, it was sufficient to use the perturbative approach
of the previous two sections. However, it is interesting to
consider the equations of motion
~(\ref{effriedmann}), (\ref{hbarprime}) and (\ref{efscalar}) in the
Einstein frame  in generality. Making the following definitions 
\bea
\tau&=&\sqrt{\rho_{cr,0}}\,\,\bar t\\ \tilde a&=&\bar a/\bar a_0\\
U&=&\gamma\,\rho_{cr,0}\,f\left(\frac{\sigma}{\sigma_*}\right) 
\eea
and redefining a prime to denote a derivative with respect to $\tau$,
the equations become
\bea \tilde H^2&=&\frac{8\pi
G}3\left[\frac{\sigma'^2}2+\gamma\,f+ \frac\alpha{1+\alpha}\,
\Omega^{-1/2}\,\tilde a^{-3}\right]\\ \sigma''&=&-3\tilde
H\sigma'-\gamma\,f_{,\sigma}+
\frac\alpha{1+\alpha}\,\frac{\Omega^{-1/2}}{2\sigma_*}\, \tilde
a^{-3}\label{scalarequaz}\\ \tilde H'&=&-4\pi G
\left[\sigma'^2+\frac\alpha{1+\alpha}\,\Omega^{-1/2}\, \tilde
a^{-3}\right]\ .  
\eea

We focus on the case of $\lambda>\alpha/2$ in order to ensure $w_{\rm
eff}<-1$ today. As we shall see, for some potentials (for example square
or exponential ones), $\sigma (\tau)$ has almost zero first derivative
($x\ll1$) but a much larger and negative second derivative, impling
that $\sigma$ achieves a local maximum today. This in turn implies a
significant fine tuning of the initial conditions for $\sigma$.

\begin{figure}[ht]
\centering
\includegraphics[width=2.5in]{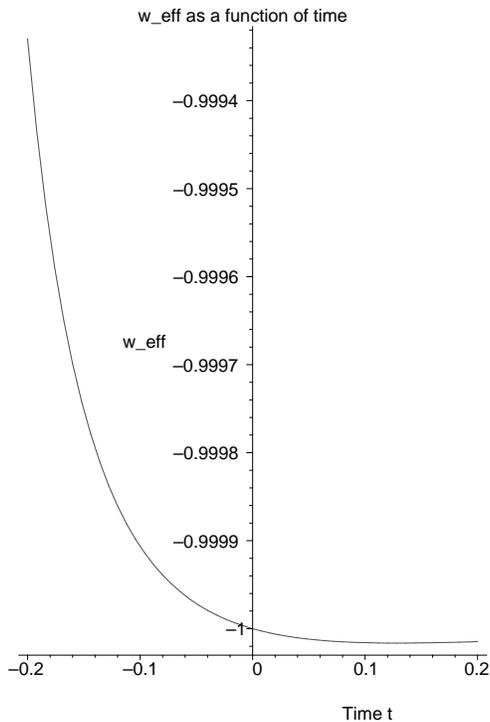}
\caption{$w_{\rm eff}$ as a function of both $y\equiv \frac1{H_0}$ and $\lambda$
with $\omega=40000$.}
\label{fig4}
\end{figure}

One possibility is to impose that $\sigma'$ is always
small, for example $\sigma'=x$ at all times, so that
$\sigma''\approx 0$, and to look for the potential that would satisfy this
condition. It is convenient to think of $f(\sigma)=f[\sigma(t)]$, so that $f$ becomes a
function of time. For example we could use the Friedmann equation, the
scalar field equation and the second Einstein equation. Since 
\be
\sqrt{\rho_{cr,0}}\left.\frac1{\tilde a}\,\frac{d\tilde a}{d\tau}\right|_0=
\left.\frac1{\tilde a}\,\frac{d\tilde a}{d\bar t}\right|_0=
\bar H_0=H_0+\frac{\sigma'_0}{2\sigma_*}\ ,
\label{eqdiva}
\ee
consistent initial conditions are
\bea
\tilde a(0)&=&1\\
\tilde a'(0)&=&\sqrt{\frac{8\pi G}3}+\frac x{2\sigma_*}\label{diva}\\
\sigma(0)&=&0\\
f(0)&=&1 \ .
\eea
We must then solve for the three new unknowns: $\tilde a$, $\sigma$, $f$. 
In this case we find that $w_{\rm eff}$ is very close to $-1$ at all
times, as shown in figure~(\ref{fig4}).

An alternative possibility is to look for a potential $f$ that leads to a
constant $w_{\rm eff}$ less than -1. This requires
\begin{equation}
\frac{\sigma''}{2\sigma_*} = \bar H'+\frac{\sigma'}{2\sigma_*}
\left(\bar H-\frac{\sigma'}{2\sigma_*}\right)+
\frac32\,(1+w_{\rm eff})\left(\bar H-\frac{\sigma'}{2\sigma_*}\right)^{\!2}
-4\pi G\,w_{\rm eff}\,\frac\alpha{1+\alpha}\,\Omega^{-3/2}\,\tilde a^{-3}\ .
\end{equation}

Using this relation, the second
Einstein equation and the scalar field equations, with initial conditions
\bea
\tilde a(0)&=&1\\
\tilde a'(0)&=&\sqrt{\frac{8\pi G}3}+\frac x{2\sigma_*}\\
\sigma(0)&=&0\\
\sigma'(0)&=&x\\
f(0)&=&1 \ ,
\eea
we find that the potential is not even a function any more ($f$ becomes 
double-valued). The potential is a
curve and $\sigma=0$ is a point at which the curve is continuous but
not differentiable. This tells us something about how unnatural a
potential like this is. So in this case a constant $w_{\rm eff}$ does not
seem likely at all.

The general issue here is the following. Since $\sigma'=x \ll 1$,
equation~(\ref{U0}) implies that $\gamma\approx 1/(1+\alpha)$. Thus, in order
to achieve $w_{\rm eff}<-1$ (e.g.~$w_{\rm eff}=-1.2$) with $\omega=1$,
equation~(\ref{lamb}) yields $\lambda\sim\alpha$.  Further,
equation~(\ref{eqdiva}) implies $\tilde H_0\approx H_0/\sqrt{\rho_{cr,0}}$ and
therefore, since $\sigma_*\,\tilde H_0\sim1$, the scalar field
equation~(\ref{scalarequaz}) evaluated at the present day is
\begin{equation}
\sigma_*\,\sigma''_0 \approx -\frac\alpha{2(1+\alpha)} \ ,
\end{equation}
where we have used $f_{,\sigma}(0)=\lambda/\sigma_*$.  This requires \be
\left|\frac{\sigma''_0\sigma_*}{\sigma'_0}\right|\gg1\quad{\rm or}
\quad\left|\frac{\sigma''_0}{\sigma'_0}\right| M_p\gg1 \ .  \ee In other
words, the function $\sigma(\tau)$ must either be
close to a maximum, having a small first derivative and a negative second
derivative or, as in section IIIA, close to a sudden change in the potential.  
Similar behavior also occurs for $\omega=40000$ with $\lambda\sim5000$.

\section{Conclusions}
The observed acceleration of the universe presents a fascinating yet daunting challenge
to particle cosmology. The development of a theoretical framework in which the correct
magnitude of the dark energy density and its relatively recent dominance are explained in 
a manner consistent with both particle physics and general relativity has so far eluded 
researchers.

Another problem is posed by the equation of state parameter of the dark energy or, 
equivalently, the time evolution of its energy density. Even for the most economical 
possibility, a cosmological constant, we are faced with a problem that has confounded 
theorists for decades. The cosmological constant problem remains even when we turn to
other dark energy models. If $-1/3 >w>-1$ then, assuming a solution to the cosmological
constant problem, there are many proposals to obtain the requisite acceleration. To date, 
however, all of these face fine tuning problems or worse, when considered as serious particle
physics or gravitational models.

A more severe problem exists in the observationally allowed range $w<-1$. Such a source
for the Einstein equations~\cite{Nilles:1983ge,Barrow:yc,Pollock:xe,Caldwell:1999ew,Sahni:1999gb,Parker:1999td,Chiba:1999ka,Boisseau:2000pr,Schulz:2001yx,Faraoni:2001tq,maor2,Onemli:2002hr,Torres:2002pe,Frampton:2002tu,Carroll:2003st} must violate the dominant energy
condition and hence may lead to instabilities of the vacuum or the propagation of energy
outside the light cone. 

One way around this problem is to consider theories in which
the Friedmann equations are not valid so that it is possible
to measure the effective value $w_{\rm eff}$ 
to be less than $-1$, even if the actual dark energy source obeys
the DEC, or if there is no dark energy at all.  

In this paper we have considered Brans-Dicke (BD) theories, to see if such 
an effect can occur naturally in this framework, while remaining consistent with other
experimental constraints such as those on the time-dependence of Newton's
gravitational constant $G$. We have demonstrated that only extremely unnatural
and contrived models lead to an inference of $w<-1$, even in this wide class
of extensions to general relativity. In particular, it is necessary to fine tune the behavior of
the Brans-Dicke scalar so that it is approaching a maximum today, having small first
derivative with respect to time and yet large second derivative.


\section*{Acknowledgments}
The authors would like to thank the KITP, Santa Barbara, where this work was initiated, for hospitality. This work was supported in part at the KITP by the National Science Foundation (NSF) under grant
PHY-9907949. The work of SC is supported in part by
U.S.~Dept.~of Energy contract DE-FG02-90ER-40560, 
NSF grant PHY-0114422 (CfCP) and the David and Lucile Packard
Foundation. The work of ADF and MT is supported in part by the NSF under grant
PHY-0094122 and by a Cottrell Scholar Award to MT from Research Corporation.


\end{document}